\documentstyle[epsf,eqsecnum,aps,multicol]{revtex}
\begin{document}
\def\thefootnote{\fnsymbol{footnote}} 
\title{Instability and front propagation in laser-tweezed lipid bilayer
tubules}
\author{Peter D.~Olmsted\footnote{e-mail: {\it
phy6pdo@irc.leeds.ac.uk}; permanent address, Dept. of Physics,
University of Leeds, Leeds LS2 9JT, UK} and F.~C.~MacKintosh\\ 
     {\small Department of Physics, University of Michigan, Ann Arbor MI
48109-2210}} 
\date{\today} 
\maketitle
\def\thefootnote{(\arabic{footnote})} 
\setcounter{footnote}{0}
\def\thesection       {\arabic{section}}
\def\thesubsection    {\arabic{section}.\arabic{subsection}}
\begin{abstract} 
We study the mechanism of the `pearling'
instability seen recently in experiments on lipid tubules under a local
applied laser intensity. We argue that the correct boundary conditions
are fixed chemical potentials, or surface tensions $\Sigma$, at the
laser spot and the reservoir in contact with the tubule. We support this
with a microscopic picture which includes the intensity profile of the
laser beam, and show how this leads to a steady-state flow of lipid
along the surface and gradients in the local lipid concentration and
surface tension (or chemical potential). This 
leads to a natural explanation for front propagation and makes 
several predictions based on the tubule length. While most
of the qualitative conclusions of previous studies remain the same,
the `ramped' control parameter (surface tension) implies several new
qualitative results.  We also explore some of the consequences of front
propagation into a noisy (due to pre-existing thermal fluctuations)
unstable medium.
\end{abstract}


\vskip 5truemm
\noindent Short Title: \ \ {\bf Dynamic instability in bilayer tubules.}

\begin{multicols}{2}
\narrowtext

\section{Introduction}
A recent series of exciting experiments \cite{barziv94} demonstrated
a dynamic instability induced on tubules of single lipid bilayers
by application of laser `tweezers', whereby the cylindrical tubule
of radius $R_0$ modulates with a wavenumber given by $q^{\ast}R_0\simeq 0.8$.
This instability has  been attributed to an excess surface tension due
to the gain in electrostatic energy when surfactant molecules, of higher
dielectric constant than water, displace water in the electric field of
the laser.

The starting point for understanding this phenomenon is the Rayleigh
instability \cite{rayleigh1,rayleigh2,tomotika} of a thin cylindrical
thread  of liquid with positive surface tension, whereby the thread can
reduce its surface area at fixed volume by modulating and evolving
towards a string of beads.  Rayleigh calculated the preferred
wavelength of a cylinder of fluid in air in the inviscid
\cite{rayleigh1} and non-inertial (viscous) \cite{rayleigh2} limits,
finding in the former case a characteristic non-zero wavenumber and in
the latter case a preferred wavenumber of zero (or infinite
wavelength).  Later, Tomotika \cite{tomotika} calculated the
instability for a viscous fluid surrounded by another viscous fluid,
again in the non-inertial regime, finding that the change in boundary
conditions restores a finite characteristic wavelength.  See Olami and Granek
\cite{granek95} for a discussion of this point.  The present problem,
however, requires a much different detailed dynamical analysis which
relates the flow of lipid molecules in the interface to the bulk flow
in the surrounding fluid. An important physical ingredient is a new
conserved quantity, the lipid on the surface.

At present there are (at least) two theoretical treatments of the
experiments of Ref.\cite{barziv94}.  Bar-Ziv and Moses \cite{barziv94} and
Nelson and co-workers \cite{nelsontube95a} have proposed the picture that
the surface tension rapidly equilibrates everywhere to an induced value
$\Sigma_0$, and the instability proceeds from this state. In contrast,
Granek and Olami \cite{granek95} have postulated that the correct
treatment of the problem is to impose a constant rate at which lipid
molecules are drawn into the trap from the tubule.  This loss of lipid is
accommodated by stretching out small wavelength surface fluctuations
and the result is again a uniform surface tension $\Sigma_0$. 
Goldstein, {\sl et al.\/} (GNPS) \cite{nelsontube95b} demonstrated
quantitatively how the equilibration of the tension in the tube stays
`ahead' of a shape change, so that a treatment with a constant (in time)
surface tension is reasonable; and argued that the primary loss of area
is in the shape instability itself, rather than through the removal of
small-scale wrinkles.

We propose a slightly different picture of the steady state before the
onset of the instability, which follows from consideration of the
experimental configuration. The tubules, as formed, are several hundred
microns long and are attached at either end to `massive lipid globules'
\cite{barziv94} of order $10\mu\hbox{m}$ in diameter. Hence, the
tubules must be in contact with a reservoir which fixes the lipid
chemical potential (or, equivalently, the surface tension). If we
assume the system is equilibrated, it follows that the chemical
potential for exchange between the tubule, reservoir, and solvent/lipid
bath vanishes \cite{schulman61}, and we may assume a reference chemical
potential of zero or, equivalently, zero surface tension. This
coincides with the experimental observation of visible thermal
fluctuations on the tubules \cite{barziv94}.

Now imagine applying a laser to the tubule.  In the electric field of
the chemical potential of a lipid molecule is lowered by an amount
$\delta\varepsilon{\cal E\/} D a$, where $D$ is the molecular length,
$\delta\varepsilon$ is the  dielectric constant relative to water, $a$
the area of the lipid, and ${\cal E\/}$ the energy density deposited in
the trap. Nelson {\sl et al.\/} \cite{nelsontube95b} calculated that
this yields an energy gain per area of bilayer of $\Sigma_0\sim 2 \cdot
10^{-3} \,\hbox{erg cm}^{-2}$, for a laser power of $50\,\hbox{mW}$.

Hence there is a large reduction in the local chemical potential as
the lipid suddenly finds it advantageous to move into the laser spot. The 
surface tension in the adjacent portion of the tubule 
increases as lipids start to move out of the surface. Since
the other end of the tubule is in contact with a reservoir at 
zero chemical potential, the final state (prohibiting, for the moment, 
surface undulations) must be a non-equilibrium
steady state in which:
\begin{enumerate}
\item Lipid is transported at constant velocity from the reservoir at
zero chemical potential to the laser trap at a negative chemical
potential.
\item The chemical potential drops linearly
along the tubule, with a gradient that balances the frictional drag
of the bulk fluid in steady state.
\item The local lipid concentration also varies linearly, since 
the two-dimensional lipid fluid membrane is compressible.
\end{enumerate}

This differs significantly from the treatments of Nelson {\sl et al.\/}
and Granek and Olami in that {\sl lipid must flow out of the anchoring
globules\/} and the chemical potential (or surface tension) never
attains a non-zero constant over the duration of the experiment. In
fact, prohibiting the shape instability, the boundary conditions
specified by both Olami and Granek {\sl and\/} Nelson {\sl et al.\/}
yield a tense final state as (a small amount of) area is drawn out of
surface fluctuations, while the treatment of the anchoring globules as
reservoirs yields the steady-state described below.\footnote{If we wait
long enough the trap will `fill up' with surfactant and the chemical
potential return to zero everywhere. However, in the present case of
strong laser power the surface instability will have occurred by this
time. See Section~3.3.}

Several consequences follow from this observation.  First, a chemical
potential gradient suggests a mechanism for front propagation
\cite{nelsontube95b,sarloos88}. The front starts at the laser spot
where the surface tension is largest, and `propagates' outward toward
the anchoring globule simply because the amplitude of the instability
grows at different rates along the tube.  Our results predict a speed of
front propagation which is inversely proportional to the length of the
tube, and is largest near the laser spot, decreasing to zero somewhere
near the anchoring reservoirs; and a characteristic wavenumber which
also decreases (much slower, see Fig.~\ref{fig:dispersion2} below) away
from the laser spot.

The outline of this paper is as follows. In Section~2 we derive the
linear concentration gradient in the absence of surface undulations.
We predict a `ramped', or spatially-varying control parameter, the
effective surface tension, which is in fact the two-dimensional
pressure whose gradient drives the flow of lipid against the viscous
drag of the bulk fluid.  In Section~3 we present a detailed microscopic
picture of the uptake of surfactant by the trap, and argue that a
competition between bending and compression energies modifies the
effective surface tension of the trap. This leads to a prediction of a
critical laser power for the onset of an instability. While this
section may safely be omitted in reading this paper, it illuminates the
nature of the instability by treating a realistic scenario for how the
trap buckles to initiate flow.

In Section~4 we discuss the implications of a slowly varying surface
tension on the detailed calculation of Goldstein, {\sl et al.\/}
\cite{nelsontube95a,nelsontube95b}.  We also discuss front propagation
within the picture of a surface tension gradient, which relates the
problem to a large body of work on front propagation with `ramped'
parameters \cite{kramer82}.  The issue of front propagation in this system
is delicate \cite{nelsontube95b}, and our results suggest at least two
possibilities, which we briefly raise in this work and pose for further
investigation.  Depending on whether noise ({\it i.e.\/} existing thermal
fluctuations in the tubule) is present, we expect front propagation
which is either (a) characteristic of that predicted by the so-called
Marginal Stability Criteria (MSC) \cite{nelsontube95b,kramer82}, or (b)
dominated by amplification of existing `noise', which can lead to behavior
reminiscent of front propagation for a steep enough ramp.  We conclude
in Section~5 by recalling the relevant timescales and frequencies, and
summarizing our predictions and the differences from previous treatments.

\section{Steady state}
In this section we calculate the steady-state configuration of a
tubule under the action of an applied laser intensity, assuming the
laser supplies a chemical potential $- \Sigma_0$ at the laser spot.
Note that this implies a reservoir in which to pack lipid molecules.
In Section~3 we support this with a microscopic picture which leads
to virtually the same results that we obtain in this section, with
a prefactor of order one which depends on the laser shape. Note that
there are several possible microscopic scenarios for initiating
flow into the trap, and Section~3 addresses only one of these.

\subsection{Equations of motion}

Changes in chemical potential $\delta\!\mu$ are related to changes in
surface tension $\delta\Sigma$ by
$\delta\mu = - \phi^{-1}\delta\Sigma $, where $\phi$ is 
the lipid concentration
(and hence $\phi^{-1}=a$ is the area per lipid). Also, 
\begin{equation}
\Sigma=-p,
\end{equation} 
where $p$ is the 2-dimensional pressure of the fluid of lipid
molecules. 

The geometry
of the system is taken as shown in Figure~\ref{fig:geom}, 
with the cylinder aligned parallel to the $z$-axis and $r$ the radial 
coordinate. The boundary conditions are
\begin{eqnarray}
p(z=0) = 0 && \qquad(\hbox{reservoir}) \\
p(z=L) = - \Sigma_0 && \qquad(\hbox{laser spot}),
\end{eqnarray} 
where $\Sigma_0$ is the surface tension induced by the laser. 
{\begin{figure}
\par\columnwidth20.5pc
\hsize\columnwidth\global\linewidth\columnwidth
\displaywidth\columnwidth
\epsfxsize=3truein
\centerline{\epsfbox{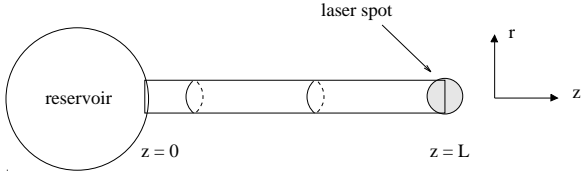}}
\caption{Geometry of lipid tubule under localized tension.}
\label{fig:geom}
\end{figure}}

The Navier-Stokes and continuity 
equations for the 2D fluid of lipid molecules are
\begin{eqnarray}
\partial_t \phi&=& -\nabla\!\cdot\!(\phi{\rm\bf v}) \label{eq:continuity}\\
\rho_s (\partial_t + {\rm\bf v}\!\cdot\!\nabla){\rm\bf v} 
&=& \eta_s \nabla^2 {\rm\bf v}
+ (\case{1}{3}\eta_s + \gamma_s) \nabla 
(\nabla\cdot{\rm\bf v})  \label{eq:NS}\\
&& - \nabla p + \Delta{\bf T}^b\!\!\cdot\!{\bf\hat{r}}. \nonumber
\end{eqnarray}
Here $\eta_s$ and $\gamma_s$ are 2D shear and bulk viscosities, $\rho_s$
is 2D lipid mass density, and $\Delta{\bf T}^b\cdot{\bf\hat{r}}$ is
the viscous drag acting on the surface from the dissipative stress tensor
$T^b_{\alpha\beta} = \case12\eta(\nabla_{\alpha}u_{\beta} +
\nabla_{\beta}u_{\alpha})$ in the surrounding fluid. This flow is 
established in a vorticity diffusion time $\tau_{v}$ which
is much smaller than other times in the problem. We ignore
drag from outside the cylinder for the moment, since this flow essentially
moves with the surfactant molecules and contributes relatively little to
the boundary stress\footnote{We may include this as, for example, the
Stokes drag on a cylinder, which increases
the right hand side of Equation~\ref{eq:stress} by of 
order 10\% \cite{landaufluids}.}. 

With the above approximation, the boundary stress is given by the shear
stress in the tube. For a 
{\sl uniform\/} flow of lipid ${\rm\bf v} = v {\bf\hat z}$, the interior flow 
is Poiseuille \cite{landaufluids},
\begin{equation}
{\rm\bf u}(r) = v {2r^2 - R_0^2\over R_0^2}{\bf\hat z},
\end{equation}
where $R_0$ is the
tube radius and we use the no-slip boundary condition $u(r=R_0)=v$. 
Hence the stress acting on the surface is
\begin{equation}
\Delta{\bf T}^b\!\cdot\!{\bf\hat{r}} = -{2\eta v \over R_0}.
\label{eq:stress}
\end{equation}

Gradients of ${\rm\bf v}$ in the $z$-direction change the flow profile from
simple Poiseuille, but this has only a very small effect on the dynamics
of establishing the steady state, primarily in the region of the laser
spot, which we ignore for now.

The final ingredient we need is the 
compressibility of the film, through the constitutive relation 
\begin{equation}
p = p_0 - \chi^{-1} \delta a 
\label{eq:pressure}
\end{equation}
where $p_0$ is the equilibrium pressure. 

Now we specialize to the problem at hand. We linearize the dynamic
equations in ${\rm\bf v}$ and $\delta\phi = \phi - \phi_0$, assume a
velocity of the form ${\rm\bf v} = v(z){\bf\hat z}$, and ignore the
inertial term in the Navier-Stokes equation. Employing Eq.~(\ref{eq:pressure}),
we obtain
\begin{eqnarray}
\partial_t \delta\phi&=& -\phi_0 \nabla_z v \label{eq:phidot}\\
0 &=& \hat{\eta} \nabla_z^2 v - B \phi_0^{-1} \nabla_z \delta \phi
+ {2\eta\over R_0} v,\label{eq:vdot}
\end{eqnarray}
where $\hat{\eta}=\case43 \eta_s + \gamma_s$.

The boundary conditions are
\begin{eqnarray}
\delta \phi(z=0)&=& 0 \\
\delta \phi(z=L)&=& \phi_0\Sigma_0/B.
\end{eqnarray} 
where $B=\chi^{-1}\phi_0^{-1}$ is the two dimensional bulk
modulus.

\subsection{Dynamics of equilibration}

Assigning $\delta\phi(z,t) = \delta\hat{\phi}(q,\omega) 
e^{i (q z - \omega t)}$ and similarly for $v(z,t)$, we obtain the
following dispersion relation:
\begin{equation}
\omega = {i B q^2 \over q^2 \hat{\eta} + 2\eta/R_0}.
\label{eq:dispersion}
\end{equation}
This yields $\omega\simeq i R_0 B q^2 / \eta$ for $q\ll q^{\ast}$
and $\omega \simeq B / \hat{\eta}$ for $q\gg q^{\ast}$, where $q^{\ast}
= \sqrt{2 \eta / (R_0 \hat{\eta})}$. Hence, at long
wavelengths we have diffusive behavior governed by the friction
against the bulk fluid, while at short wavelengths the dynamics is
dominated by the 2D viscosity. 
The crossover length is given by
$1/q^{\ast} \sim 0.1\mu\hbox{m}$, where we have taken $\eta \sim 
10^{-2} \,\hbox{g cm}^{-1}\hbox{s}^{-1}, \hat{\eta} \sim 10^{-6} \,\hbox{g
s}^{-1}$, and $R_0\sim 0.5 \mu\hbox{m}$. Hence in most cases of interest
we are in the regime dominated by bulk fluid dissipation and may
ignore $\hat{\eta}$.

We can now estimate (within linear response) the time to attain steady state
after imposing the localized tension by the laser as, roughly, the
relaxation time of the slowest mode given by the dispersion relation
Eq.~(\ref{eq:dispersion}). Taking $q = 2\pi/L$, we have
\begin{equation}
\tau_{\mit ss} \sim {2 L^2\eta\over (2\pi)^2 R_0 B}\sim 10^{-5}\,\hbox{s},
\end{equation}
for $B\sim 150 \,\hbox{erg cm}^{-2}$ \cite{evans87} and $L\sim 100\mu\hbox{m}$.
This estimate of $B$ is a zero-temperature estimate and ignores 
small-scale thermal fluctuations which soften this modulus considerably
\cite{helfrich84,nelsontube95b}. As we discuss in the conclusion and as
shown in Reference~\cite{nelsontube95b}, 
this effect can reduce $B$ by up to three orders of magnitude, increasing 
$\tau_{\mit ss}$ accordingly, to of order $10^{-2}\,\hbox{s}$.
We can compare this to the vorticity diffusion time,
\begin{equation}
\tau_{\it v} = {\rho R_0^2 \over \eta} \sim  10^{-7}\hbox{s},
\end{equation}
where we take $\rho=1\,\hbox{g cm}^{-3}$. Since $\tau_{\mit ss}>\tau_v$
our assumption above of a uniform shear stress is reasonable.

\subsection{Steady State}

To find the steady state we equate the left hand side of
Eq.~(\ref{eq:phidot}) to zero, which yields a constant velocity
$\bar{v}$. From Eq~(\ref{eq:vdot}) $\nabla_z \delta \phi$
is also a constant and, applying the boundary conditions and the
pressure constitutive equation, Eq.~(\ref{eq:pressure}), we 
find the following steady-state profile 
\begin{eqnarray}
\bar{v} &=& { R_0 \Sigma_0\over 2 \eta L} \label{eq:vbar}\\
\delta \phi &=& -\phi_0 {\Sigma_0 \over B} {z\over L} \\
\delta\Sigma &=& \Sigma_0 {z\over L} \label{eq:sigma}.
\end{eqnarray} 

Thus we find that the steady state, excluding modulations of the cylinder,
is a non-equilibrium steady state, where the lipid molecules run down a
chemical potential gradient and the molecular spacing increases to reflect
this changing local potential.  An estimate above yields $\bar{v} \sim
1 \mu\hbox{m s}^{-1}$, where we use $\Sigma_0 \sim 10^{-3} \hbox{erg
cm}^{-2}$. The effective surface tension (or two dimensional pressure) is
induced by the applied laser, and is non-zero {\it only in the presence
of flow\/}.

\section{Microscopic Picture}
We have shown how lipid flow, which is a necessary condition for an
effective non-zero surface tension far from the trap, follows from a
boundary condition of fixed chemical potential at the trap. In this
section we argue that this boundary condition requires an instability
in the trap, and we present detailed calculations for a possible
scenario for the trap to initiate flow.  We stress that there are
several possible mechanisms, including buckling, ejection of micelles
or bilayer structures, and growth of `cancerous' membranes.  This is
surely not an exhaustive list.

\subsection{Basic Considerations}
We first note that a laser spot centered at $z=L$ typically has a Gaussian
intensity profile \cite{webb81}, which leads to an energy gain per area
of lipid ${\cal U\/}(z) = - \Sigma_0\,\zeta(z)$, with \begin{equation}
\zeta(z) = e^{-(z-L)^2/2\Delta^2}.  \end{equation} The spot radius was
estimated to be $\Delta\simeq 0.15 \mu\hbox{m}$ in the experiments of
Bar-Ziv and Moses \cite{barziv94}.

We can envision two scenarios after applying the laser: 
\begin{itemize}
\item[(a)] Lipid can be sucked into the trap
until the electrostatic energy gain balances the
cost of compressing the molecules in the bilayer. At this point
the trap is full, flow stops, and the chemical potential (and
surface tension) of the entire tube reverts back to zero.
\item[(b)] For a critical tension
$\Sigma^{\ast}$ (Eq.~\ref{eq:deltabar0}) we expect the compressed section 
of the tubule to become unstable with respect to buckling.
For higher intensities the trap continues to fold to accommodate more
lipid, initiating a flow along the tubule. This flow must be
accompanied by a chemical potential (or surface tension) gradient, 
which drives the instability seen in the experiments. 
\end{itemize}

Our discussion suggests that the trap boundary condition should contain the
physics that, at a certain distance from the center of the trap, lipid
is incorporated into folds to relieve the in-layer compression. A
reasonable choice is
\begin{equation}
a(z=L-\bar{\Delta})=a_0, \label{eq:BCtrap}
\end{equation}
where $\bar{\Delta}$ is a distance to be determined. 
This asserts that the area per head assumes its preferred equilibrium 
value at the point where the folding begins.

In the next two subsections we derive the steady-state flow into the
trap (prohibiting for the moment the `pearling' shape change). We first
obtain a general relation for the local area per lipid 
$a(z)$, which depends on the trap boundary condition. We then deduce
a crude criterion for the position $z=L-\bar{\Delta}$ at which the trap
buckles. Applying the {\sl assumption\/} of Eq.~(\ref{eq:BCtrap}) at this
position then yields the desired profile and steady-state flow. 

\subsection{Detailed Steady State}
Let us examine the steady state. The continuity
equation, Eq.~(\ref{eq:continuity}), yields the condition
\begin{equation}
v(z) = C a(z),
\end{equation}
where $C$ is a constant to be determined.
Hence the steady-state
Navier-Stokes equation, ignoring the 2D viscosity, becomes
\begin{equation} 
0 = {2\eta C a(z) \over R} + \chi^{-1} \nabla_z a(z) - \nabla {\cal
U\/}(z).
\end{equation} 

{\begin{figure}
\par\columnwidth20.5pc
\hsize\columnwidth\global\linewidth\columnwidth
\displaywidth\columnwidth
\epsfxsize=3truein
\centerline{\epsfbox[18 110 552 482]{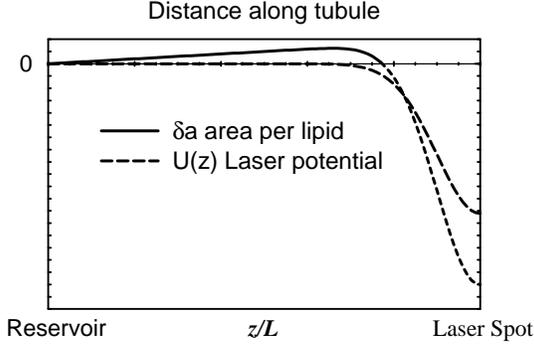}}
\caption{Profiles for steady-state deviation $\delta a$ in the area per lipid
and the potential set up by the laser after buckling has occurred.
The dotted line shows the continuation of $\delta a$
(Eq.~\protect{\ref{eq:solution}}) into the buckled trap region beyond 
the point $z=L-\bar{\Delta}$, and should actually be replaced with
buckled membrane at $\delta a=0$. The boundary conditions are: (1)
$\delta a=0$ at the reservoir, due to fixed chemical potential; and
(2) $\delta a(z=L-\Delta)=0$ (see Eq.~\protect{\ref{eq:BCtrap}}). 
In steady state $\delta a$
increases linearly away from the reservoir because the
membrane is under tension, which varies linearly to counteract 
the viscous drag due to lipid flow.  }
\label{fig:profiles}
\end{figure}}

The solution of this equation with the boundary condition $a(0)=a_0$,
which follows from contact with a reservoir at ambient pressure, is
\begin{eqnarray}
{\delta a \over a_0} &=& e^{\lambda z} - 1 - \gamma\int_0^z\!ds\,
e^{-\lambda(s - z)} {\partial \over\partial s} \zeta(s) 
\label{eq:solution}\\
\lambda &=& {2\eta\chi C\over R_0}, \\
\gamma &=& {\Sigma_0\over B},
\end{eqnarray} 
where $\lambda$ is determined by the boundary condition at the trap.
Note that $\gamma \sim 10^{-5}$, since 
$\Sigma_0\sim 10^{-3} \,\hbox{erg cm}^{-2}$ and 
$B\sim 150 \,\hbox{erg cm}^{-2}$. 
Using Eq.~(\ref{eq:BCtrap}), we find 
\begin{equation}
\lambda= {- \gamma \,\zeta(L-\bar{\Delta})\over L - \bar{\Delta} -
\int_0^{L-\bar{\Delta}}\!ds\, \zeta(s)},
\end{equation}
where we have expanded for small $\lambda$ and made use of $\zeta(0)=0$ far
from the trap. For all practical purposes $\lambda= - 
\gamma \,\zeta(L-\bar{\Delta})/L$.

Since $\lambda L \ll 1$ for most cases of experimental interest,
the calculation of Section~2 applies in the region
outside the trap (see Fig.~\ref{fig:profiles}) with the 
surface tension replaced by 
\begin{equation}
\Sigma_0 \rightarrow \bar{\Sigma} = \Sigma_0\,\zeta(L-\bar{\Delta}).
\label{eq:barsig}
\end{equation}
For a Gaussian shape $\zeta(L-\bar{\Delta}) \alt 1$ 
[$\zeta(L-\bar{\Delta})\simeq 0.31$ in Fig.~\ref{fig:buckle}], 
so the modification to the naive boundary condition is rather
minimal.

\subsection{Trap Boundary Condition}
To complete our discussion we estimate the stability against folding
inside the trap. This determines the position $\bar{\Delta}$ at which
the boundary condition (\ref{eq:BCtrap}) applies, as well as a critical
effective surface tension parameter (or laser intensity) $\Sigma^{\ast}$
at which the system initiates flow.  We imagine that the system has
attained a steady state in the absence of buckling and flow, given
by Eq.~(\ref{eq:solution}) with $\lambda=0$: \begin{equation} a(z) =
a_0\left[1 - \gamma\,\zeta\left(z\right)\right].  \end{equation} Here
$\gamma\,\zeta\left(z\right)$ is a measure of the compression.

Rather than calculating the stability of a patch with a non-uniform area
per head $a(z)$, we
calculate the stability against buckling of a patch with uniform $a=\psi a_0$,
with $\psi= 1 - \epsilon $, and use the resulting critical strain 
$\epsilon^{\ast}$  to 
determine $\bar{\Delta}$ through
\begin{equation}
\gamma\,\zeta(L-\bar{\Delta}) \equiv \epsilon^{\ast}.
\label{eq:deltabar}
\end{equation}

We consider perturbations $R(z) = R_0(1 + u(z))$ which preserve the
volume of the fluid. This constraint yields the condition 
$\int [u(z)^2 + 2 R_0 u(z)] =0$ \cite{safran}. 
The free energy, which includes in-plane
compression and bending, is \footnote{We do not include a term involving
interaction with the laser, since we are interested in an instability at
fixed particle number on the membrane.}
\begin{equation}
2 F = \int d^2\!r \left[ B \left({\psi a - a_0 \over a_0}\right)^2 
+ \kappa H^2 \right],
\end{equation}
where $H$ is the mean curvature, $a_0 = dz R_0$, and $a =  \psi d^2\!r$.
For the perturbation above \cite{granek95},
\begin{eqnarray}
d^2\!r  &=&  dz (R_0 + u(z)) \sqrt{1 + u'(z)^2} \\
H  &=&  {(1 + u'(z)^2)^{-1/2}\over R_0 + u(z)} - {u''(z)\over(1 +
u'(z)^2)^{3/2}},
\end{eqnarray} 
where $u'(z) = du/dz$. 

To quadratic order in $u(z) = \sum_q \hat{u}(q) e^{iqz}$ the energy
per unit area $A$ becomes \cite{granek95}
\begin{eqnarray}
{2 F\over A} &=& \sum_q \hat{u}(q)^2 \left[ {3\kappa\over R_0} - 5 B
R_0 \epsilon\right.  \nonumber\\
&& - \hat{q}^2 \left.\left(3B R_0 \epsilon + {\kappa\over R_0}\right)
+ 2 \hat{q}^4 {\kappa\over R_0}\right],
\end{eqnarray}
where $\hat{q}=q R_0$.
The vanishing of the term in square brackets defines 
$\epsilon^{\ast}$, the minimum strain above 
which this energy is unstable to undulations.
Combining this condition with our estimate for how $\epsilon(z)$ varies
away from the trap, Eq.~(\ref{eq:deltabar}), we find the following
relation which determines $\bar{\Delta}$:
\begin{eqnarray} 
\sigma^{\ast}\equiv{\Sigma^{\ast} R_{0}^2 \over \kappa}  &=& 
{1\over \zeta(L-\bar{\Delta})} {3 + 2 \bar{q}^4 -
\bar{q}^2 \over 5 + 3 \bar{q}^2} \label{eq:deltabar0}\\
&\equiv& g(\bar{q})
\end{eqnarray} 
where $\bar{q} \simeq \pi R_0/\bar{\Delta}$. For $\sigma>\sigma^{\ast}$
the tubule should buckle inside the trap. Fig.~\ref{fig:buckle} shows
$g(\bar{q})$ for the Gaussian laser spot. 

{\begin{figure}
\par\columnwidth20.5pc
\hsize\columnwidth\global\linewidth\columnwidth
\displaywidth\columnwidth
\epsfxsize=3truein
\centerline{\epsfbox{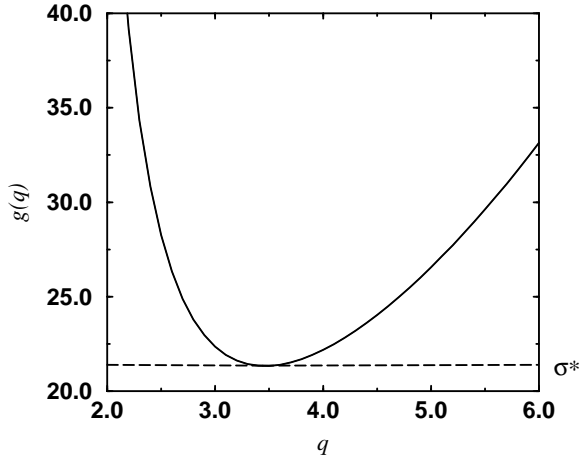}}
\caption{Buckling criterion inside the laser trap for a laser spot size
$\Delta=0.15\,\mu\hbox{m}$ and tubule radius $R_0=0.5\,\mu\hbox{m}$.
Here, we find $\sigma^{\ast}\simeq 21, \bar{\Delta} \simeq 0.9 R_0$, 
$\zeta(L-\bar{\Delta})\simeq 0.31$, and $\bar{\Sigma} \simeq 0.97 \Sigma_0$.}
\label{fig:buckle}
\end{figure}}

Our criterion depends on the trap shape, but not the compression
modulus $B$. This happens because, while the strain induced in packing
lipid in the trap varies as $1/B$, the critical strain at which
buckling occurs is also inversely proportional to $B$, and the $B$
dependence cancels out. In fact, the same order of magnitude estimate
emerges from a comparison of bending and effective surface tension. What
we have gained, however, is a picture of the forces at play that induce
the buckling.

While our estimate apparently fails to predict buckling for typical
values, relaxing a few approximations we have made should change this
picture. First, we have assumed uniform coverage at the lowest value of
the actual nonuniform coverage in the quiescent trap. Second, we have
assumed an axisymmetric deformation. This is obviously not the case if
the laser spot size is smaller that the tubule diameter. In addition,
the volume constraint must be handled differently. Removing these
approximations should, in both cases, result in a smaller $\Sigma^{\ast}$.
For example, in the limit of small trap sizes we can ignore curvature
and ask about the stability of a flat interface against buckling, for
which the criterion above becomes
\begin{equation}
\Sigma^{\ast} = {\kappa \bar{q}^2\over\zeta(L-\bar{\Delta})}.
\end{equation}
demanding that the characteristic buckling wavevector $\bar{q}$ be
roughly the inverse of the trap size $\bar{\Delta}$ determines the
critical trap size and intensity. This is essentially the same estimate
given by Bar-Ziv, Frisch, and Moses in a somewhat different context
\cite{barziv95}.

\section{Dispersion Relation and Front Propagation}
\subsection{Dispersion Relation}

We have shown thus far that, under steady-state conditions before any
macroscopic shape instability occurs, the proper boundary conditions
imply a surface tension gradient along the tubule which supports lipid
flow. We now turn to the effects of this
gradient. Rather than repeating the the analysis of
GNPS \cite{nelsontube95b} with a non-uniform surface tension, we note
that the characteristic wavenumber at which the instability occurs
is typically $q^{\ast}R_0\simeq 0.8$. Since $R_0\ll L$, we suspect
that the assumption of a locally constant surface tension along the
tubule is a good first step. This allows us to transcribe the results
of Ref.~\cite{nelsontube95b}.

The primary result of interest is the growth rate $\omega(q)$ of an undulation
$u(q,t)$, where $q$ denotes a Fourier mode along the tubule. This 
frequency is defined through 
\begin{equation}
({\partial\over \partial t} + i q \bar{v}) u(q,t) = \omega(q) u(q,t),
\end{equation} 
where the convective term arises because the lipids have an average
velocity. 

In the original instability calculation presented by Rayleigh
\cite{rayleigh1}, and as has been emphasized in
Refs.\cite{barziv94,nelsontube95a,granek95,nelsontube95b}, the structure
of $\omega(q)$ is as follows:
\begin{equation}
\omega(q) = \Phi(q) T(qR_0).
\end{equation}
The function $T(qR_0)$ is determined by the energetics of the problem and,
in our case, is non-zero for $\Sigma R_0^2/\kappa$ greater than a
critical value of order unity \cite{nelsontube95b,granek95}. In the
Rayleigh case the instability occurs for $\Sigma>0$. The function
$\Phi(q)$ is determined by the dynamics of the problem, and it is here
that much of the interesting and surprising physics lies. Energetics
tells us that the most unstable modes are at low $q$, where undulations
are the least `violent', while dynamic considerations severely
penalize the growth of modes in the limit $q\rightarrow 0$.

Goldstein {\sl et al.\/} \cite{nelsontube95b} 
calculated $\omega(q)$ for a uniform surface
tension $\bar{\Sigma}$, including the effects of bending as well as
friction between the two bilayer leaves.  Changing the boundary conditions
of their work to allow for flux from the reservoir adds a convective term
to the dynamics, and, aside from our local approximation above, changes nothing
else. A plot of $\omega(q)$ is shown in Figure~\ref{fig:dispersion}
for several values of the surface tension $\sigma =\bar{\Sigma} R_0^2
/ \kappa$, with values for the bilayer friction and bulk modulus taken
as in GNPS.

{\begin{figure}
\par\columnwidth20.5pc
\hsize\columnwidth\global\linewidth\columnwidth
\displaywidth\columnwidth
\epsfxsize=3truein
\centerline{\epsfbox{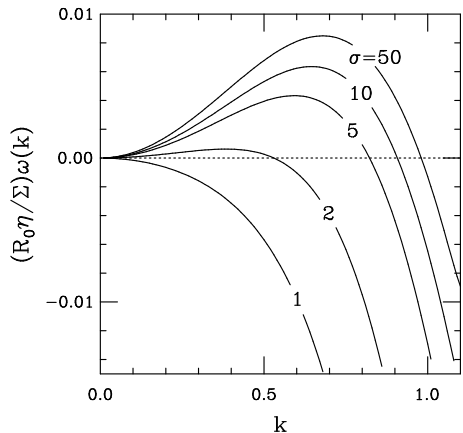}}
\caption{Dispersion relation $\omega(k)$, with parameters
as calculated by GNPS, 
$\beta=3.5, \epsilon=0.5$. Reprinted from 
Ref.~[7]. Alternatively, we may consider
this as a plot of $\omega(k,z)$ for  $\sigma=50$ and $z=L, L/5,
L/10, L/25,$ and $L/50$. Here, $k$ is in units of $2\pi/R_0$.}
\label{fig:dispersion}
\end{figure}}

Since $\sigma$ is $z$-dependent (Eq.~\ref{eq:sigma}), 
the growth rate $\omega^{\ast}$  and 
wavenumber $q^{\ast}$ of the fastest
growing mode are $z$-dependent, and are greatest near
the laser spot, as in Fig.~\ref{fig:dispersion2}. 
A single Fourier modulation
has, locally, the following form: \cite{nelsontube95b}
\begin{equation}
u(z,t) = u_0 e^{i q (z - \bar{v} t) + \omega(q,z)t}, 
\end{equation}
where $\omega(q,z)$ is the function plotted as $\omega(q)$ in
Fig.~5 of Reference~\cite{nelsontube95b} and is reproduced here in
Fig.~\ref{fig:dispersion}. The $z$-dependence comes through the
$z$-dependence of $\sigma$. Note that we rely
{\sl strongly\/} on the condition $q L \gg 1$ [Note that in the
experiments \cite{barziv94} with, say, $L\simeq 200\,\mu\hbox{m}$ and a 
diameter of $1\,\mu\hbox{m}$, this condition is easily satisfied, 
$qL\simeq 160$.]

Given a dispersion relation which depends on position, there are several
immediate naive predictions: The local wavenumber and apparent growth
rate of the pattern 
should decrease as the anchoring globules are approached, with
the instability vanishing at a point close to the reservoir
where the induced surface tension is not strong enough to overcome the
barrier due to bending, $\Sigma^{\ast} \simeq \kappa/R_0^2$. Hence, in the
experiments with $\sigma\simeq 20$ \cite{nelsontube95a}, this occurs at
$1/20th$ of the distance from the anchoring globule
to the laser trap. At points closer to the globule any undulation is a
decaying remnant of the pattern developed closer to the trap.
{\begin{figure}
\par\columnwidth20.5pc
\hsize\columnwidth\global\linewidth\columnwidth
\displaywidth\columnwidth
\epsfxsize=3truein
\centerline{\epsfbox{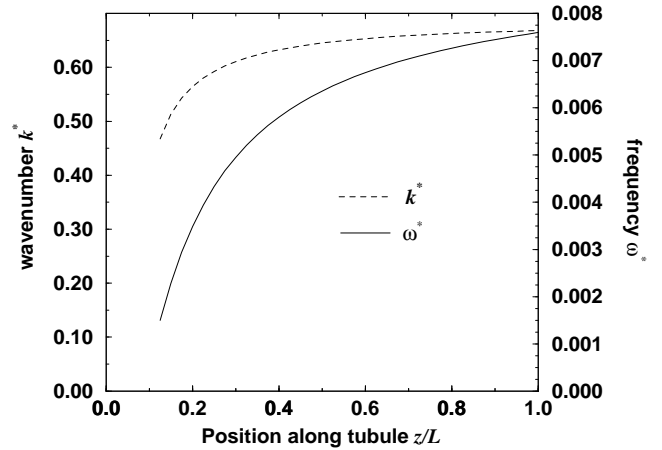}}
\caption{Frequency $\omega^{\ast}$ and dimensionless wavenumber 
$k^{\ast}$ of fastest
growing mode as a function of position along the tubule, obtained from
Fig.~\protect{\ref{fig:dispersion}} by taking $\sigma(z=L)=20$.}
\label{fig:dispersion2}
\end{figure}}

\subsection{Front Propagation}

We now confront the issues of front propagation and wavelength
selection. GNPS argued that the Marginal Stability Criterion hypothesis
provides a reasonable estimate for both the propagation speed $v_f$ and
the selected wavenumber $k^{\ast}$ \cite{nelsontube95b}. A naive
extension of this calculation, again assuming a local dispersion
relation, predicts a spatially varying front speed and selected
wavenumber. However, this prediction and the qualitative picture relies
on two assumptions: (1) the existence of a propagating front, and (2)
the absence of noise---i.e., that the propagation occurs into a
uniform, unstable medium with no thermal fluctuations.  The latter
assumption is obviously not correct in detail, as thermal fluctuations,
including modes of wavelengths comparable to the most unstable modes,
are apparent in the experiments prior to the onset of the instability.
Here we discuss these issues in the context of a spatially varying
control parameter.  Because a full treatment of the problem does not
yet exist in the literature and is beyond the scope of this paper, we
limit ourselves in this work to some numerical experiments and
suggestions which, we hope, will stimulate further research on both
this specific problem and the general aspects of front propagation into
spatially-varying media in the presence of noise. For the rest of this
paper we refer to `noise' as a set of random initial conditions which
obey a Boltzmann distribution, and do not consider temporally
fluctuating noise.

Kramer {\sl et al.\/} \cite{kramer82}, followed by others
\cite{buell86,kramer85,riecke87}, showed that, in the presence of a
`ramped' control parameter that becomes subcritical at some point, as
happens near the reservoir, the uniquely-selected wavenumber need not
correspond to that determined by the MSC.  This effect is expected to
take precedence over the MSC-determined wavenumber at times after which
non-linearities become important and the `phase' of the pattern has time
to diffuse of order the system size. However, we are concerned with the
more fundamental issue of the {\it existence\/}  of a propagating front.

In the presence of `noise' which, in the present experiment,
corresponds to existing thermal fluctuations around the reference
smooth cylindrical state, a propagating front can be expected to exist
for times less than the characteristic growth times of existing
fluctuations in the vicinity of the most unstable mode. Hence, given a
quench into an unstable `ramped' state with an initial perturbation
near the laser spot, propagation away from the perturbation occurs for
an initial period of time, followed by rapid growth all along the
cylinder as the initial conditions (`noise' of unstable wavelengths) are
amplified to visible length scales. An initial perturbation near the
laser spot is natural because, in practice, the laser spot diameter is
smaller than the tubule diameter and a `pinching' effect results
whereby surfactant flows around the circumference of the tubule (as
well as along the cylinder diameter) to fill the trap.

The effect of a `ramp' in the control parameter should be most dramatic
after the noise overwhelms the front propagation: for a flat control
parameter (no ramp) the noise grows randomly everywhere, and the
`front' should break down when the noise has grown to visible
amplitudes. However, for a steep enough ramp the non-uniform
amplification of the noise could resemble front propagation.

To check these conjectures we have 
employed a simple caricature of the tubule dynamics, 
specified by
\begin{equation}	
({\partial\over \partial t} + \bar{v} \partial_x) u(x,t) = 
\left[a(x) 2 k_0^2 \partial_x^2  + \partial_x^4 \right]
u(x,t) - g u(x,t)^3, 
\label{eq:caricature}
\end{equation} 
where $a(x)$ is a spatially varying control parameter chosen to mimic
the dispersion relation and position dependencies in
Figs.~\ref{fig:dispersion} and~\ref{fig:dispersion2}.
A choice which gives reasonable qualitative agreement is
\begin{equation}
a(x) = {|x-x_0|^{\alpha+1}\over (x-x_0)},
\label{eq:trial}
\end{equation}
where $x_0$ is the point at which the system is absolutely unstable.
We emphasize that this is a toy model whose details do not correspond to
the Bar-Ziv {\sl et al.\/} experiments, but which we believe contains the 
essential physics of front propagation into an unstable inhomogeneous medium,
as occurs in these experiments.
For Fig.~\ref{fig:dispersion2}, $x_0\sim 0.05 L$ and $\alpha=1/8$ are
reasonable. Fig.~\ref{fig:mimic} shows the local dispersion relation
$\omega^{\ast}(x)$ for various $\alpha$.
{\begin{figure}
\par\columnwidth20.5pc
\hsize\columnwidth\global\linewidth\columnwidth
\displaywidth\columnwidth
\epsfxsize=3truein
\centerline{\epsfbox{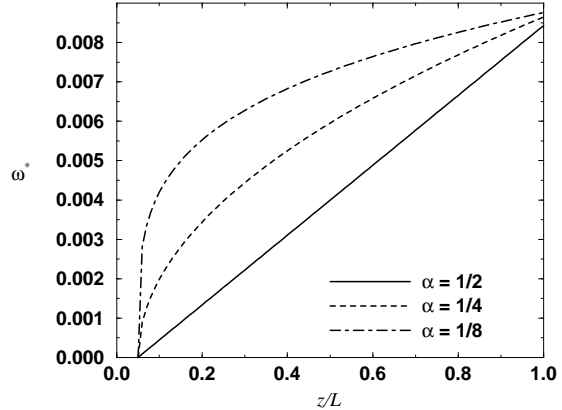}}
\caption{Local dispersion relation $\omega^{\ast}$ for trial form,
Eq.~(\protect{\ref{eq:trial}}), for $\alpha=1/2,1/4,1/8$. The trap is at
$z/L=1$.}
\label{fig:mimic}
\end{figure}}

We have chosen the simplest possible non-linearity to stabilize the
system. One may choose a more physical non-linearity such
as the driving force arising from terms of higher than quadratic order
in the mean curvature \cite{nelsontube95b}, but our purpose here
is primarily to illustrate some qualitative behavior of a front
propagating into a noisy, non-uniform media.

Figs.~\ref{fig:snapshots1}~and~\ref{fig:snapshots2} show snapshots in
the evolution of the system, given by Eq.~(\ref{eq:caricature}), for
an initial perturbation at the trap of $1\%$ of the final amplitude
(as determined by the non-linear term $g$) and an initial condition
(or `noise') which is taken to be a superposition of $300$ harmonics
weighted with a Boltzmann weight corresponding to a non-zero surface
tension ({\it i.e.\/} with an energy proportional to $q^2$). We have
chosen a system size of $150$ wavelengths, and arbitrarily chosen the
vertical scale to fill the figures.

The general features are as described above: a front `propagates' for
an initial time from the initial perturbation, after which the `noise'
takes over and a very irregular growth quickly overtakes the system. 
{\begin{figure}
\par\columnwidth20.5pc
\hsize\columnwidth\global\linewidth\columnwidth
\displaywidth\columnwidth
\epsfxsize=3truein
\centerline{\epsfbox{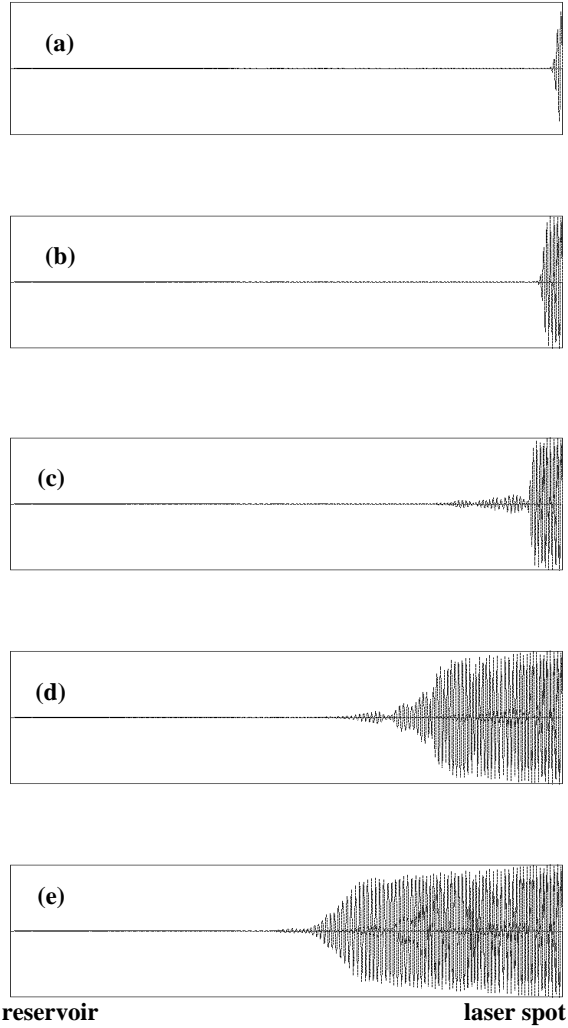}}
\caption{Consecutive snapshots of $u(x,t)$ as determined by 
Eq.~(\protect{\ref{eq:caricature}}). The conditions are: convection
$\bar{v}=0$; initial 
perturbation at the laser spot $ u(L,t=0) = 10^{-1}u(L,\infty)$;
ramp parameter $\alpha=1/2$;
initial mean noise amplitude a fraction 
$10^{-6}$ of the final amplitude. Time intervals are every $80$ time steps.
The vertical scale is chosen to fit the amplitude of the undulation, and as
such is a different scale than the horizontal scale. All snapshots have
the same vertical scale. (a) and (b) show a propagating front; (c) and
(d) show the acceleration due to amplification of the noise, and (e) shows a
return to apparent propagation.}
\label{fig:snapshots1}
\end{figure}}
The steeper ramp ($\alpha=1/2$, Fig.~\ref{fig:snapshots1})
has a better defined growth in the `noise' regime, and could almost
be called a `front'. In contrast, the growth into the shallow ramp
($\alpha=1/8$, Fig.~\ref{fig:snapshots2}) is more ragged and it
would be charitable to call this a front. The shallower ramp has a very
slightly faster propagation speed in the initial regime, and an
obviously faster `propagation' speed in the noise-dominated regime. Both
of these behaviors may be traced to the faster overall growth rate for a
shallower ramp (where a larger fraction of the tubule is more unstable).

{\begin{figure}
\par\columnwidth20.5pc
\hsize\columnwidth\global\linewidth\columnwidth
\displaywidth\columnwidth
\epsfxsize=3truein
\centerline{\epsfbox{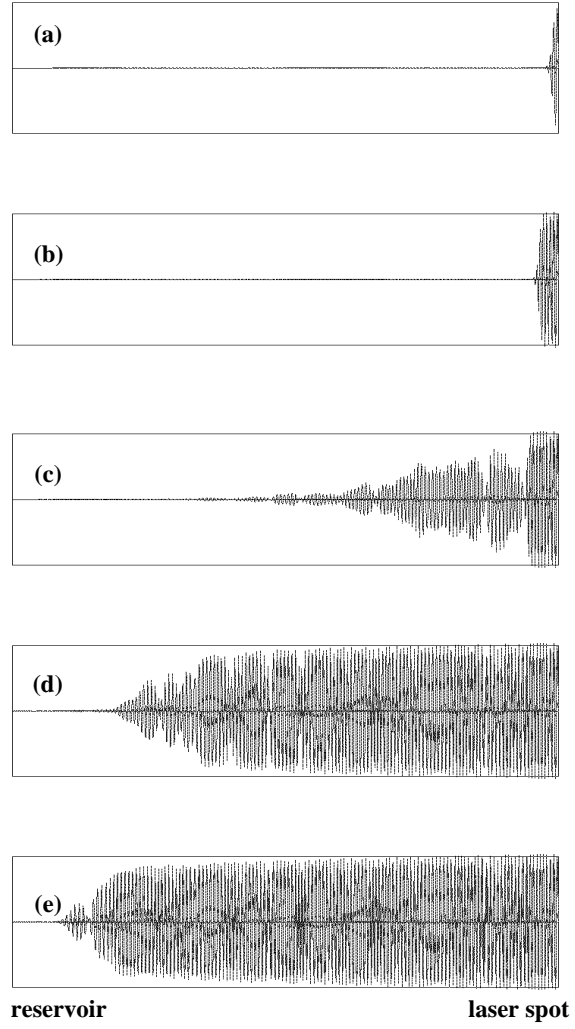}}
\caption{Same parameters and time steps as in
Fig.~\protect{\ref{fig:snapshots1}}, 
except with ramp parameter $\alpha=1/8$.}
\label{fig:snapshots2}
\end{figure}}

Fig.~\ref{fig:noise} shows the the results of fixing the ramp and
varying the noise amplitude.  For a noisier system the effective
propagation speed in the noise-dominated regime is faster, and the
breakdown of the simple propagating regime occurs earlier. The
propagation velocity before the noise takes over is independent of the
noise amplitude. The delay before noise-dominance increases
logarithmically with increasing noise amplitude, consistent with the
simple argument that the propagating solution exists until the noise
has grown to a given amplitude, since this initial growth is
exponential.

To summarize, we have performed exploratory numerical calculations to
investigate some of the consequences of a ramped control parameter,
with an initial localized perturbation and initial global `noise' for
an initial condition, finding: 
\begin{enumerate}
\item At early times a
front propagates away from the localized perturbation. We find a
dimensionless front velocity of $v\,\omega(k_0)/k_0 = 3.3$, while the
Marginal Stability
Criteria \cite{sarloos88} predict $4.6$. A similar agreement was found 
in the simulations of GNPS \cite{nelsontube95b}.
{\begin{figure}
\par\columnwidth20.5pc
\hsize\columnwidth\global\linewidth\columnwidth
\displaywidth\columnwidth
\epsfxsize=3truein
\centerline{\epsfbox{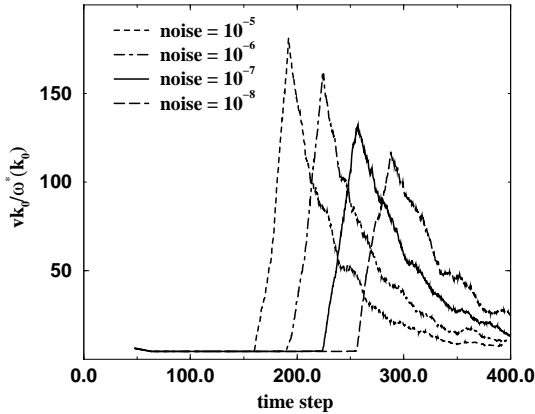}}
\caption{Front velocity vs. time, averaged over $100$ realizations of
initial noise with weights described in the text. The velocity was
measured by tracking the leading edge of the envelope of wavelets,
and is plotted in units of the
characteristic velocity $\omega(k_0)/k_0$, where $\omega(k)=2\,k_0^2\,
k^2 - k^4$ is the dispersion relation for Eq.~(\protect{\ref{eq:caricature}})
for $a(x)=1$. Ramp parameter $\alpha=1/8$. The noise is given as 
a fraction of the final amplitude.}
\label{fig:noise}
\end{figure}}

\item In this first regime the propagation velocity is independent of
noise amplitude.
\item After a time, which may be taken to be the time required for the
`noise' to grow to an observable amplitude, the unstable pattern rapidly
develops everywhere on the tubule. 
\item The speed and qualitative character of this growth depend on the
noise and ramp characteristics. The growth is faster for a shallower
ramp and/or stronger noise, and looks reminiscent of a front for a
steeper ramp; and in all cases is much faster than the simple front
propagation from the initial localized perturbation at the laser trap.
\end{enumerate}

We have also performed calculations with no initial localized
perturbation but, as this is probably not physically relevant, we do
not report the results here.  This initial study raises several questions
which we feel are worth pursuing. In cases where the noise is weak
enough and an apparent front exists, can this be understood
quantitatively in terms of the gradient in the control parameter, and
how does this relate to previous investigations of `ramped' control
parameters \cite{kramer82,buell86,kramer85,riecke87}? Can a ramp
stabilize an advancing front?

\section{Conclusion}
\subsection{Physical Picture}

We have given the following picture of the action of lipid tubules upon
the application of laser tweezers. In the absence of buckling, the
laser induces a local compression of lipid molecules in the laser spot.
This takes place in a time of order $\tau_{\mit ss}\sim 10^{-5}
\hbox{s}$. A sufficiently large laser intensity induces a local
buckling of the membrane in the trap, which initiates flow down the
tubule from the reservoir. We do not have an estimate of the delay time
for the instability in the trap. In the absence of undulations {\sl
outside\/} the trap, this flow would build up to a steady-state value
$\bar{v}\sim 1\,\mu\hbox{m s}^{-1}$ in a time order $\tau_{\mit ss}$.
The physics of this flow is a balance between drag against the bulk
fluid and a force due to the gradient imposed by the chemical potential
drop between the reservoir and the trap.

Given the steady-state tension profile within the membrane and the
reasonable assumption that the gradient occurs over a length (L) much
larger than the critical wavelength ($\sim R_0$), the analysis of GNPS
\cite{nelsontube95b} leads to a Rayleigh-like instability to undulations.
This instability initiates near the laser spot where the chemical
potential (or surface tension) is lowest, and propagates away from the
spot to a point along the tubule at which the local surface tension falls
below the critical tension $\Sigma_{cr}=\kappa/R_0^2$ which characterizes
the instability. Typical growth frequencies are $\omega\sim 25\,\hbox{s}^{-1}$
\cite{nelsontube95b}, which corresponds to times $\tau_{\omega}\sim
10^{-1}\hbox{s}$. We note that the experiments find a significant time
delay of order seconds before the instability, \cite{mosesprivate},
a still-unexplained observation.

The estimate above for $\tau_{\mit ss}$ assumes a bare $2D$ compression
modulus $B$, while GNPS (see \cite{helfrich84}) pointed out that $B$
undergoes significant softening at the lengthscale of the tubule, due to
the thermal fluctuations at low surface tension, and estimated a decrease
of up to three orders of magnitude. This, correspondingly, would increase
the value of $\tau_{\mit ss}$ to $10^{-2}\,\hbox{s}$, so that  an accurate
quantitative calculation must include the dynamics of the increase in
surface tension. This leads to, effectively, a smaller applied tension
$\sigma$ and hence a slower propagation speed.  This spirit was followed
in the approach of Granek and Olami \cite{granek95}.

Our interpretation assumes that the trap accommodates material by
folding, or some other means. Our analysis suggests that the proper
boundary condition should be a fixed surface tension
$\bar{\Sigma}<\Sigma_0$ at the laser spot, where the laser shape
determines $\bar{\Sigma}$ through Eq.~(\ref{eq:barsig}) and
$\bar{\Delta}$ through Eqs.(\ref{eq:deltabar},\ref{eq:deltabar0}).

This reservoir picture suggests that, upon
turning off the laser, the system can revert to the original
tubule by unfolding or, if severe topological changes have
occurred (by, for example, budding in the laser spot or the creation of
metastable `pearls' as seen in the experiments), attain some other
long-lived metastable state. 

\subsection{Discussion}

In this work we have made several assumptions. The assumption that we can
treat the anchoring globules as reservoirs presupposes that any damping
processes retarding the transfer of lipid to and from the globules is
negligible relative to other dynamical processes. We expect this to
arise from the same source as the two-dimensional surface viscosity,
which we have argued in Section~2.2 to be negligible. We have given
a simplified picture of the scenario of trap buckling, where we take
a single characteristic buckling wavevector, and treat the trap as
uniform. This `single-mode' approximation may be naive, and preliminary
calculations suggest that the system is in fact less stable than this
simple analysis would suggest \cite{pdofcmunpub}. There are also several
other possible modes of instability which we have only mentioned but
which could certainly play a role. We have also specified a boundary
condition at the trap whereby the the lipid relaxes to its preferred area
per head group, which seems reasonable but is not otherwise justified.
Finally, we have made a local approximation for the variation of the
surface tension so that we may use the results of GNPS. This applies
for sufficiently long tubules, $L/R_0\gg 1$.

Front propagation and the detailed effects of
propagating into a spatially-varying medium have only been touched
upon in our numerical treatment. This study still leaves much to
be resolved; one important question is how to accurately treat the
non-linear regime.  This has been treated in different ways by
Olami and Granek \cite{granek95} who considered the non-linear effect
of removing surfactant from the membrane in the {\sl absence\/} of a
gradient, and by Goldstein {\sl et al.\/} \cite{nelsontube95b}, who added
the correct non-linear terms in the bending energy to examine the
propagation of the pearls.

The primary new ingredients in our theory are (1) our treatment of the
anchoring lipid globules as reservoirs and (2) our exploratory treatment
of the role of pre-existing thermal fluctuations (noise) in determining
the `front-like' characteristics of the instability.
Both Nelson and co-workers
\cite{nelsontube95a,nelsontube95b} and Olami and Granek \cite{granek95}
`turn off' the reservoir. In the latter case material is drawn out of the
existing thermal fluctuations, while Nelson and co-workers attribute the
area change primarily to the shape instability itself. Olami and Granek
impose a constant flux boundary condition at the trap, while Nelson
and co-workers impose a fixed chemical potential $-\bar{\Sigma}$ at the
trap which, fairly rapidly, reduces the chemical potential everywhere
to $-\bar{\Sigma}$.  Our picture essentially gives the same boundary
condition at the trap, but the treatment of the globule as a reservoir
changes the qualitative picture dramatically.

Our theory differs from previous theories in several respects, and
there are many consequences which may be checked experimentally.
Obviously, we expect flow when an instability develops. This could be
visualized by, for example, fluorescence spectroscopy with a very
dilute fraction of labelled lipids. The inhomogeneous surface tension
implies that the local dispersion relation is also spatially-dependent,
as in Fig~\ref{fig:dispersion2}, which implies that the velocity of
front propagation $v_f$ (which is proportional to $\omega^{\ast}$
\cite{nelsontube95b}) and characteristic wavenumbers should decrease
farther away from the laser spot. Note that the characteristic
wavelength changes very gradually compared to the speed of propagation,
and as such would be more difficult to detect. It would also be
interesting to see, experimentally, whether fluctuations are actually
strong enough to destroy the front-like character, or whether two
characteristic regimes exist in the experiments, as indicated in
Fig.~\ref{fig:noise}.  Finally, we mention that the opportunity of
using laser pulses to control flow within lipid and other systems
presents amusing possibilities and applications.

\acknowledgements
It is a pleasure to thank E.~Moses, R.~Granek, P.~Nelson, T.~Powers, 
C.-M.~Chen, S.~Milner, and W. van~Saarloos for helpful conversations and 
correspondence. This work was
supported in part by NSF Grant No.~DMR 92-57544 and by The Donors of the
Petroleum Research Fund, administered by the American Chemical Society.

\end{multicols}

\begin{thebibliography}{10}

\bibitem{barziv94}
R. Bar-Ziv and E. Moses, {\sl Instability and ``pearling''states produced in
  tubular membranes by competition of curvature and tension}, Phys. Rev. Lett.
  {\bf 73} (1994) 1392.

\bibitem{rayleigh1}
{Lord~Rayleigh}, {\sl On the instability of jets}, Proc. Lond. Math.~Soc. {\bf
  10} (1879) 4.

\bibitem{rayleigh2}
{Lord~Rayleigh}, {\sl On the instability of a cylinder of viscous liquid under
  capillary force}, Phil.~Mag. {\bf 34} (1892) 145.

\bibitem{tomotika}
S. Tomotika, {\sl On the instability of a cylindrical thread of a viscous
  liquid surrounded by another viscous fluid}, Proc. Roy.~Soc. Lond. {\bf A150}
  (1932) 322.

\bibitem{granek95}
R. Granek and Z. Olami, {\sl Dynamics of Rayleigh-like instability induced by
  laser tweezers in tubular vesicles of self-assembled membranes}, J.~Phys.~II
  (France) {\bf 5} (1995) 1349.

\bibitem{nelsontube95a}
P. Nelson, T. Powers, and U. Seifert, {\sl Dynamical theory of the pearling
  instability in cylindrical vesicles}, Phys. Rev. Lett. {\bf 74} (1995) 3384.

\bibitem{nelsontube95b}
R.~E. Goldstein, P. Nelson, T. Powers, and U. Seifert, {\sl Front propagation
  in the pearling instability of tubular vesicles}, J.~Phys.~II (France) {\bf
  6} (1996) 767.

\bibitem{schulman61}
J.~H. Schulman and J.~B. Montagne, {\sl Formation of microemulsions by amino
  alkyl alcohols}, Ann.~N.Y. Acad. Sci. {\bf 92} (1961) 366.

\bibitem{sarloos88}
W. van Sarloos, {\sl Front propagation into unstable states}, Phys.~Rev. {\bf
  A37} (1988) 211.

\bibitem{kramer82}
L. Kramer, E. Ben-Jacob, H. Brand, and M.~C. Cross, {\sl Wavelength selection
  in systems far from equilibrium}, Phys. Rev. Lett. {\bf 49} (1982) 1891.

\bibitem{landaufluids}
L.~D. Landau and E.~M. Lifschitz, {\em Fluid Mechanics} (Pergamon, Oxford,
  1959).

\bibitem{evans87}
E. Evans and D. Needham, {\sl Physical properties of surfactant bilayer
  membranes: thermal transitions; elasticity; rigidity; cohesion; and colloidal
  interactions}, J.~Phys. Chem. {\bf 91} (1987) 4219.

\bibitem{helfrich84}
W. Helfrich and R.-M. Servuss, {\sl Undulations steric interaction and cohesion
  of fluid membranes}, Nuovo Cim. {\bf 3D} (1984) 137.

\bibitem{webb81}
M.~B. Schneider and W.~W. Webb, {\sl Measurement of submicron laser beam
  radii}, Appl. Opt. {\bf 20} (1981) 1382.

\bibitem{safran}
S.~A. Safran, {\em Statistical Thermodynamics of Surfaces, Interfaces, and
  Membranes} (Addison-Wesley, Reading, MA, 1994).

\bibitem{barziv95}
R. Bar-Ziv, T. Frisch, and E. Moses, {\sl Entropic expulsion in vesicles},
  Phys. Rev. Lett. {\bf 75} (1995) 3481.

\bibitem{buell86}
J.~B. Buell and I. Catton, {\sl Wavenumber selection in ramped
  Rayleigh-{B}\'enard convection}, J.~Fluid Mech. {\bf 171} (1986) 477.

\bibitem{kramer85}
L. Kramer and H. Riecke, {\sl Wavelength selection in Rayleigh-{B}\'enard
  convection}, Z.~Phys.~B - Cond.~Matt. {\bf 59} (1985) 245.

\bibitem{riecke87}
H. Riecke and H.-G. Paap, {\sl Perfect wave-number selection and drifting
  patterns in ramped Taylor vortex flow}, Phys. Rev. Lett. {\bf 59} (1987)
  2570.

\bibitem{mosesprivate}
E.~Moses, private communucation.

\bibitem{pdofcmunpub}
P.~D.~Olmsted and F.~C. MacKintosh, unpublished (1996).

\end{thebibliography}
\end{document}